\documentclass[lettersize,journal]{IEEEtran}
\usepackage{amsmath,amsfonts}
\usepackage{algorithmic}
\usepackage{algorithm}
\usepackage{array}
\usepackage[caption=false,font=scriptsize,labelfont=sf,textfont=sf]{subfig}
\usepackage{textcomp}
\usepackage{stfloats}
\usepackage{url}
\usepackage{verbatim}
\usepackage{graphicx}
\usepackage{cite}
\usepackage{color}
\usepackage{multirow}
\def \etal {\emph{et al.}}

\begin{document}

\title{MSP-Conversation: A Corpus for Naturalistic, Time-Continuous Emotion Recognition}

\author{Luz~Martinez-Lucas~\IEEEmembership{Student-Member,~IEEE,}  
        Pravin~Mote,~\IEEEmembership{Student~Member,~IEEE,}     
        Abinay~Reddy~Naini,~\IEEEmembership{Student~Member,~IEEE,}
        Mohammed~Abdelwahab,
        and Carlos~Busso,~\IEEEmembership{Fellow,~IEEE}
\IEEEcompsocitemizethanks{
\IEEEcompsocthanksitem This work was supported by the National Science Foundation (NSF) under Grants CNS-1823166 and CNS-2016719. 

\IEEEcompsocthanksitem L. Martinez-Lucas, P. Mote, and A. Reddy~Naini are with the Erik Jonsson School of Engineering and Computer Science, University of Texas at Dallas, Richardson, TX 75080 USA and the Language Technologies Institute, Carnegie Mellon University, Pittsburgh PA-15213 USA (luz.martinez-lucas@utdallas.edu, AbinayReddy.Naini@utdallas.edu, Pravin.Mote@UTDallas.edu).
\IEEEcompsocthanksitem C. Abdelwahab is with AT\&T Labs Research, Bedminster, NJ USA (ma069q@att.com).
\IEEEcompsocthanksitem C. Busso is with the Language Technologies Institute, Carnegie Mellon University, Pittsburgh PA-15213 USA (busso@cmu.edu).}
\thanks{Manuscript received March 11, 2026; revised ?}}

\maketitle
\thispagestyle{plain}
\pagestyle{plain}

\begin{abstract}
Affective computing aims to understand and model human emotions for computational systems. Within this field, \emph{speech emotion recognition} (SER) focuses on predicting emotions conveyed through speech. While early SER systems relied on limited datasets and traditional machine learning models, recent deep learning approaches demand large-scale, naturalistic emotional corpora. To address this need, we introduce the MSP-Conversation corpus: a dataset of more than 70 hours of conversational audio with time-continuous emotional annotations and detailed speaker diarizations. The time-continuous annotations capture the dynamic and context-dependent nature of emotional expression. The annotations in the corpus include fine-grained temporal traces of valence, arousal, and dominance. The audio data is sourced from publicly available podcasts and overlaps with a subset of the isolated speaking turns in the MSP-Podcast corpus to facilitate direct comparisons between annotation methods (i.e., in-context versus out-of-context annotations). The paper outlines the development of the corpus, annotation methodology, analyses of the annotations, and baseline SER experiments, establishing the MSP-Conversation corpus as a valuable resource for advancing research in dynamic SER in naturalistic settings.
\end{abstract}


\section{Introduction}
\label{sec:Intro}

\IEEEPARstart{A}{ffective} computing is a field that focuses on the development of computational approaches to model, analyze, understand, recognize, and synthesize aﬀective behaviors \cite{Picard_2000}. This field is particularly relevant to applications in human-computer interactions, where fully understanding what a person is communicating is important. Human communication often involves understanding the emotions of others. Misunderstanding or ignoring people's emotions can lead to inefficient interactions. Given the ubiquity of speech-based devices, it is appealing to build systems able to recognize emotions communicated through speech. The field of \emph{speech emotion recognition} (SER) focuses on developing computing systems that can recognize the emotionality of human speech \cite{Busso_2013}. Modern SER systems, built with deep learning-based models, require a large amount of data with recordings that resemble the spontaneous externalization of emotion during daily interactions \cite{Wani_2021}. This paper describes the MSP-Conversation corpus, a speech dataset of natural emotional conversations. The corpus contains 310 conversations (77 hrs 26 mins) of natural interactions, providing a great resource to advance modern SER-based systems. 

\IEEEpubidadjcol

SER systems recognize emotion at specific temporal levels, including sentence, utterance, word, and phone. Previous SER studies have predominantly focused on identifying the emotion of an utterance or speaking turn \cite{Swain_2018}. The common formulation in SER is to recognize the global emotional label assigned to an utterance. This formulation builds on many popular emotional speech databases, which contain utterance-level global annotations, assigning a single emotional value to a full speech utterance \cite{Burkhardt_2005,Busso_2008_5,Grimm_2008,Upadhyay_2023_2,Busso_2025}. An alternative strategy for annotating emotional data is with time-continuous annotations, where emotional traces describing an emotional attribute are provided for each frame/sample \cite{McKeown_2012,Ringeval_2013,Kossaifi_202x}. These time-continuous annotations have some advantages over utterance-level annotations: (1) annotations done by different raters necessarily have the same context. It is important for raters to share similar context during the annotation process since previous work has shown that what raters annotate before their current ratings affects their emotional evaluations \cite{Martinez-Lucas_2023,Martinez-Lucas_202x}. (2) Speech utterances can have subtle to explicit emotional changes within them, and utterance-level annotations cannot capture these within-utterance dynamics. In time-continuous annotations, raters can capture all emotional dynamics. (3) Utterance-level annotations allow raters to contemplate on their emotional evaluations of singular utterances, which is not how people typically evaluate emotions in natural interactions. Time-continuous annotations contain instantaneous rater reactions, which result in important differences in annotation values \cite{Martinez-Lucas_2024_2}. (4) Annotations can be aggregated into flexible temporal levels, facilitating utterance, phase, syllable, and phone-based analysis. Given these advantages, we collect the MSP-Conversation corpus which contains longer speech recordings of human interaction that are annotated with time-contiguous labels. The human annotations are obtained using a joystick to record the instantaneous emotional perceptions of the raters while they listen to each conversation.

The speech data in the MSP-Conversation corpus is sourced from podcasts used in the collection of the MSP-Podcast corpus \cite{Busso_2025}. These podcasts come from publicly available audio sources and are chosen for their clean and emotional speech. The MSP-Podcast corpus consists of isolated speaking turns annotated with global, out-of-context utterance-level labels. Therefore, it is not appropriate to study contextual information during natural conversations. The purpose of the MSP-Conversation corpus is to complement the MSP-Podcast corpus with time-continuous annotations, providing an ideal resource for understanding the dynamic expression of emotions in human interaction. The MSP-Conversation corpus intentionally includes audio segments that overlap with the MSP-Podcast corpus, enabling direct comparisons between time-continuous annotations and sentence-level labels for valence, arousal, and dominance \cite{Martinez-Lucas_2024_2} (Sec. \ref{ssec:msp-pod_overlap}). There are 12,555 speaking turns in the MSP-Podcast corpus that are also in the MSP-Conversation corpus.

During the development of the MSP-Conversation corpus, we focused on four key features:
\begin{enumerate}
    \item Natural speech, i.e., spontaneous speech with no control or prompts from researchers.
    \item A variety of speakers with a focus on multi-party conversations.
    \item A variety of topics and conversation structures, e.g., interviews, talk-shows, lectures, reviews.
    \item Large amount of speech data to train modern SER models.
\end{enumerate}
Many datasets that contain emotional speech and time-continuous annotations have one or more of the outlined features. To the best of our knowledge, no other dataset contains all four of these features (Sec. \ref{sec:RelWork}), which limits their use in a variety of applications. The MSP-Conversation corpus can be used in applications that involve both structured and unstructured speech with an undefined number of speakers. 

We further analyze the annotations of the corpus by plotting the distribution of the labeled emotional values and calculating inter-evaluator agreements. The inter-evaluator agreements for the full corpus show that our labels are reliable and that most of our evaluators increase agreements when they are included (Sec. \ref{ssec:agreements}).  The overlap with the MSP-Podcast corpus allows us to use the information of  speaker identities and meta-data from the MSP-Podcast corpus to create speaker-independent data partitions (Sec. \ref{ssec:data_sets}), which we provide in our release. The multi-party conversations in the MSP-Conversation corpus are important but complex data. Therefore, we also conduct manual speaker diarizations of all the conversations (Sec. \ref{sec:SpkDiarizations}). The statistics from the diarizations show that the majority of the corpus includes multi-party interactions and that there is a balance between apparent gender of the speakers.

The paper is structured as follows: Section \ref{sec:RelWork} discusses other time-continuous emotional datasets that include speech data, highlighting the additional features introduced by our corpus. Section \ref{sec:DataCollection} explains how we chose the speech data for the corpus. Section \ref{sec:Annotation} explains the annotation process, analyzes the emotional content of the corpus, and evaluates the annotations using inter-evaluator agreement. Section \ref{sec:partitions} describes the pre-defined data partitions recommended for training SER models and compares our time-continuous annotations with the MSP-Podcast annotations. Section \ref{sec:SpkDiarizations} explains how the speaker diarizations were collected and provides statistics from these diarizations. Section \ref{sec:SER} presents baseline evaluations for dynamic SER experiments using multiple machine learning methods. Section \ref{sec:discussion} discusses research paths that are enabled by the MSP-Conversation corpus. Finally, Section \ref{sec:conclusions} concludes the paper.

\section{Related Work}
\label{sec:RelWork}

\begin{table}[!t]
\centering
\caption{Related datasets with time-continuous emotional traces and speech data.}
\begin{tabular}{@{}l@{\hspace{0.1cm}}c@{\hspace{0.1cm}}c@{\hspace{0.1cm}}c@{\hspace{0.1cm}}c@{\hspace{0.1cm}}c@{\hspace{0.1cm}}c@{}}
    \hline
    \textbf{Dataset} & \textbf{Size} & \textbf{\#spk} & \textbf{Avail} & \textbf{Type} & \textbf{Size} & \textbf{\# Spkr} \\
    \hline
    CreativeIT \cite{Metallinou_2016} & $\times$ & $\times$ & \checkmark & Acted & $\sim$ 8 hrs & 16 \\
    SEMAINE \cite{McKeown_2012} & $\times$ & $\times$ & \checkmark & Natural & 15.83 hrs & 28 \\
    RECOLA \cite{Ringeval_2013} & $\times$ & $\times$ & \checkmark & Natural & 3.83 hrs & 46 \\
    AVEC'14 \cite{Valstar_2014} & $\times$ & $\times$ & \checkmark & Natural & -- & 84 \\
    Aff-Wild2 \cite{Kollias_2018} & \checkmark & \checkmark & \checkmark & Natural & $\sim$ 60 hrs & 460 \\
    SEND \cite{Ong_2019} & $\times$ & $\times$ & \checkmark & Natural & 7.25 hrs & 49 \\
    SEWA DB \cite{Kossaifi_202x} & \checkmark & \checkmark & \checkmark & Natural & $\sim$ 33 hrs & 398 \\
    MuSe-CaR \cite{Stappen_2021} & \checkmark & $\times$ & \checkmark & Natural & 36.87 hrs & 90 \\
    DynAMoS \cite{Girard_2023} & $\times$ & $\times$ & \checkmark & Acted & 1.53 hrs & 22 \\
    \hline
    MSP-Conversation & \checkmark & \checkmark & \checkmark & Natural & 77.45 hrs & $>$450 \\
    \hline
\end{tabular}
\vspace{0.1cm}
\label{tab:other_datasets}
\end{table}

\subsection{Related Datasets}

Several emotional corpora have been developed to study affective behavior across diverse contexts. They vary in terms of spontaneity, annotation schemes, and recording setups, providing complementary perspectives on how emotion is expressed and perceived in speech. This section focuses on emotional databases using time-continuous annotations, similar to the MSP-Conversation corpus. Table \ref{tab:other_datasets} summarizes the corpora reviewed in this section and illustrates the distinguishing features of the MSP-Conversation corpus relative to earlier databases.

Early emotional datasets annotated with time-continuous annotations were recorded in controlled environments. For the USC CreativeIT database \cite{Metallinou_2010_2, Metallinou_2016}, actors engaged in goal-driven improvisational interactions, recorded with full-body motion capture, audio, and video, designed to elicit expressive speech and body language. The SEMAINE corpus \cite{McKeown_2012} captured emotionally rich, audiovisual interactions between human participants and \emph{sensitive artificial listener} (SAL) agents. Each session involved a user conversing with a trained operator who adopted a fixed emotional persona, such as cheerful or gloomy. Although the SAL agents were able to elicit some natural affective responses, the interactions remained semi-structured rather than entirely spontaneous. In the RECOLA corpus \cite{Ringeval_2013}, pairs of French-speaking participants collaborated on a survival task via video conferencing, with mood induction procedures used to elicit different affective states. Participant's physiological signals, such as \emph{electrocardiogram} (ECG) and \emph{electrodermal activity} (EDA), were also recorded. Although more natural than the SEMAINE corpus, RECOLA still relied on mood induction and a structured task. In the \emph{audio/visual emotion challenge} (AVEC) 2014 corpus \cite{Valstar_2014}, participants engaged in short task-oriented dialogues while being recorded with synchronized audio and video. Although less controlled, the AVEC 2014 interactions are still derived from constrained interview-style scenarios. These early structured data collection environments limited the naturalness, diversity, and size of the datasets. The MSP-Conversation corpus builds on existing recordings, enabling more natural interactions and scalable expansion across a wide range of sources.

More recent datasets have focused on recording data in naturalistic environments, resulting in more spontaneous speech. The \emph{Stanford emotional narratives dataset} (SEND) \cite{Ong_2019} has audiovisual videos of authentic, self-paced storytellings by participants sharing personal emotional experiences. Instead of scripted dialogues, listeners engaged with unscripted narratives, resulting in natural narratives. However, the collection process still limits the diversity and size of the dataset. Moreover, the data only contains monologues, as participants tell stories rather than engage in multi-party dialogue. Monologues can differ substantially from multi-person conversations in the emotions they elicit \cite{Chien_2025,Chien_2022}, making SEND unfit for use in interactive applications. The \emph{SEWA database} (SEWA DB) \cite{Kossaifi_202x} is an in-the-wild audiovisual dataset intended for emotion and sentiment modeling across culturally diverse participants. The corpus has two tasks for the participants: (1) watching commercial clips, and (2) discussing the content of these clips on video chat. The participants were recorded in unconstrained real-world settings via webcam and microphone. SEWA DB surpasses many earlier corpora by offering a broad range of cultural backgrounds and spontaneous interactions. Still, the SEWA DB data collection process limits the emotional behaviors of its data. A structured process cannot capture the full range of interaction types, as in the data-collection process used in the MSP-Conversation corpus.

Many datasets have adopted data collection practices similar to those used in the MSP-Conversation corpus to increase data quantity and diversity. The Aff-Wild2 dataset \cite{Kollias_2018} contains in-the-wild data from 558 YouTube videos, capturing a wide range of facial expressions, head poses, and lighting conditions. The corpus has become a central benchmark for studying emotion recognition under unconstrained real-world conditions. However, the dataset is primarily visual and focused on single-person recordings. Similarly, the MuSe-CaR dataset \cite{Stappen_2021} opens up the study of sentiment and affect in everyday, user-generated car review videos by capturing over 36 hours of multimodal data from YouTube, including speech, video, and text. Its focus also remains on monologue-style recordings rather than interactive conversations. Both the Aff-Wild2 and the MuSe-CaR datasets cannot be used to model interactions between speakers, making them unfit for various applications. The MSP-Conversation corpus comprises a range of spontaneous single-speaker and multi-party recordings, making it suitable for a variety of applications. Another recent database that takes its data from existing recordings is the DynAMoS database \cite{Girard_2023}. It features 22 carefully selected movie clips in English and utilizes 80 raters for emotional annotations. The main feature of the dataset is its focus on inter-rater differences for more descriptive and naturalistic annotations. The goal was to increase the number of annotators, even with a limited number of videos. In contrast, the MSP-Conversation corpus focuses on naturalistic, diverse, and multi-party speech data ideal for building SER models.

\subsection{Emotional Descriptors}

Studies on SER have primarily focused on describing emotions with emotional categories, such as happiness, sadness, and anger \cite{Swain_2018}. While these categories are natural ways to describe emotions, they have drawbacks: natural emotions are complex and nuanced \cite{Kehrein_2002,Mower_2009}. Emotional categories cannot capture this complexity without significantly increasing the number of classes \cite{Cowie_2003} and accounting for the co-occurrence of emotions \cite{Chou_2022_2}. Emotional attributes describe the emotion of an event as a point in a multi-dimensional space, with axes that describe emotional properties. This alternative representation can alleviate some of these problems.  In the MSP-Conversation corpus, we utilize the emotional attributes of valence (negative to positive), arousal (calm to active), and dominance (weak to strong). These emotional attributes were introduced in the core affect theory, \cite{Russell_1999,Russell_2003}, and can effectively characterize nuanced emotions \cite{Fontaine_2007}.

\section{Data Collection Process}
\label{sec:DataCollection}

Our goal is to have conversations among multiple people that convey natural emotions. Recording people in the wild is expensive and time-consuming. Therefore, we focus on using existing speech data available in online repositories. The MSP-Podcast corpus \cite{Lotfian_2019_3,Busso_2025} contains speech sentences that are extracted from publicly available podcasts. These podcasts were selected to include a diverse set of emotions across a broad range of topics. We also extract our conversations from these podcasts.

The MSP-Podcast corpus was annotated out of order, with global labels representing the emotion of the speaking turns. The MSP-Podcast corpus comprises only selected segments with durations between three and eleven seconds. These speaking turns were automatically retrieved using emotion recognition models. Therefore, the MSP-Podcast corpus is not ideal for studying the dynamics of emotion in conversation, which motivated the collection of the MSP-Conversation corpus as a companion dataset. The MSP-Conversation comprises 10-20-minute segments with manually annotated speaking turns and emotions, providing a valuable resource to complement the MSP-Podcast corpus.

We pick each podcast and conversation in three steps: (1) we look at the MSP-Podcast sentences in each podcast and choose the ones where the annotated sentences show extreme emotions or high emotional variability. As the corpus grew, we also focused on balancing the emotional content of the dataset as characterized with the attributes of valence, arousal and dominance. (2) We manually listen to each podcast and further filter by topic and content. 
(3) We manually choose start and end times for one conversation in each podcast. The goal is for the conversation to have a natural start and end while encompassing many of the desired emotional sentences. We pick conversations between 10 and 20 minutes long.

Annotating these long conversations in a time-continuous manner without a break is mentally taxing. To reduce cognitive load for the workers, we further split each conversation into 3- to 7-minute segments. These splits are done manually at natural breaks in each conversation. We call these 3 to 7 minute-long segments \emph{conversation parts}.

\section{Annotation Process}
\label{sec:Annotation}

\begin{figure}[t]
    \centering
    \includegraphics[width=\linewidth]{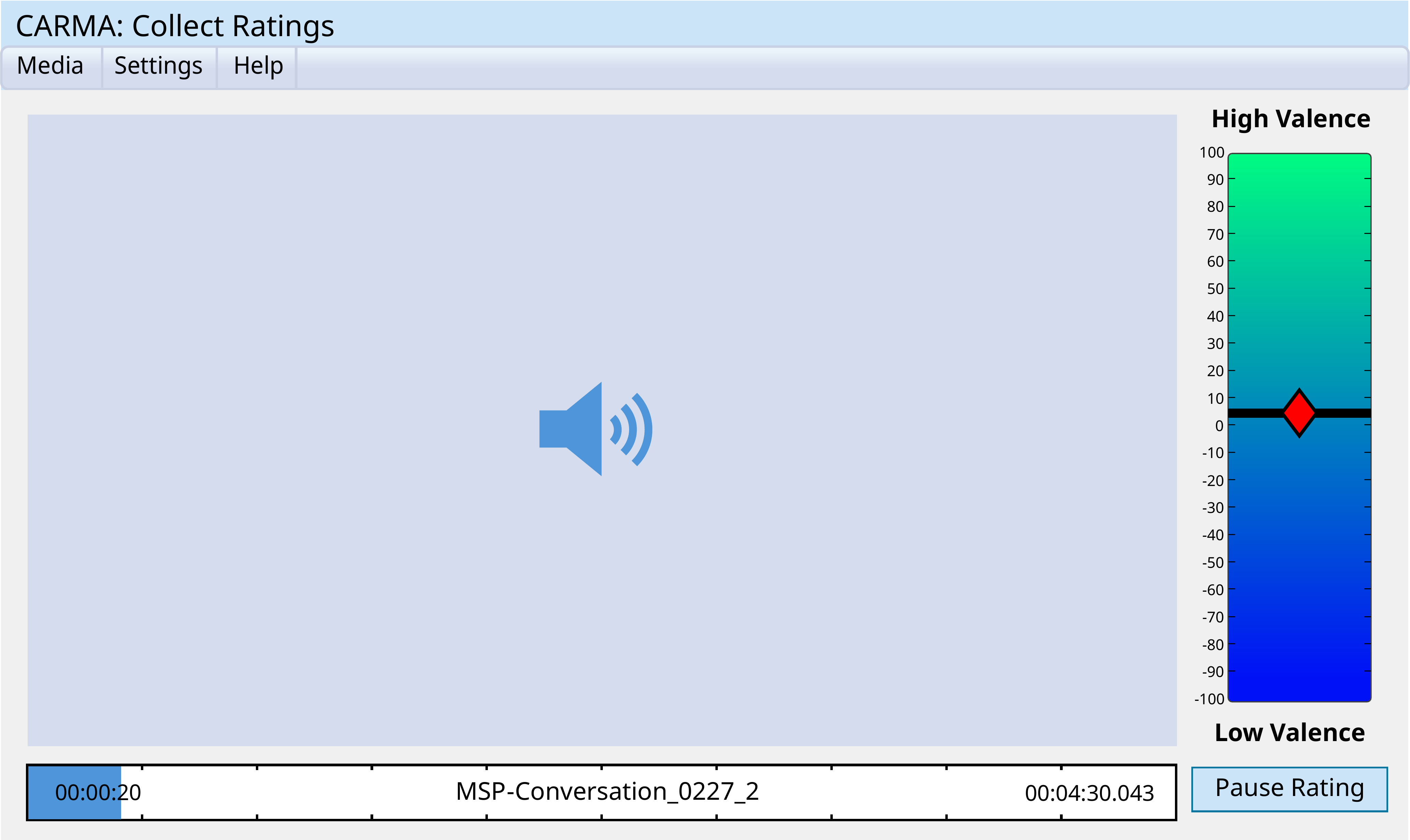}
    \caption{Graphic user interface of CARMA during annotation of the MSP-Conversation corpus. The example shows a valence annotation of the \emph{MSP-Conversation\_0227\_2} conversation part.}
    \label{fig:carma}
\end{figure}

The corpus annotation is performed using the MATLAB-based CARMA \cite{Girard_2014} software, which is designed for time-continuous annotations. The workers use a joystick to record their emotional perception as they listen to each conversation part. The workers conducted annotations for valence (negative to positive), arousal (calm to active), and dominance (weak to strong) one at a time. Figure \ref{fig:carma} shows the \emph{graphic user interface} (GUI) used during the annotation of
valence for the conversation part named \emph{MSP-Conversation\_0227\_2}. The right section of Figure \ref{fig:carma} shows the track where the position of the joystick is shown. The extreme values on that track are displayed as \emph{High} and \emph{Low} valence (the GUI displays valence, arousal, or dominance depending on the target attribute being annotated). The center points of the track correspond to the neutral values of the attributes. The joystick position is mapped to the range -100 to 100 and represents the perceived attribute value at that moment. An advantage of using a joystick for these annotations is that the workers receive tactile feedback when they reach the extremes or central values during annotation. The joystick cannot go further when it reaches the extremes, and it snaps back to the central position when it is let go.

As mentioned in the previous section, emotional annotations are done on the conversation parts of the corpus. The first step is recruiting workers. For a given attribute and conversation part, our goal is to obtain annotations from at least six different workers.  Emotional perception depends on cultural background. Therefore, we focus on workers who have lived extensively in the United States, as they are familiar with its cultural norms. We recruited workers from the \emph{University of Texas at Dallas} (UT Dallas) student body. We created a job posting only available to students and alumni of UT Dallas. After interviewing the selected applicants, we selected the workers based on availability and knowledge of emotional concepts. The recruitment of applicants was done multiple times throughout the development of the corpus. The full corpus contains annotations from 26 workers.

After recruiting workers, a researcher met with each of them to loan out a joystick, set up CARMA on their preferred system, and train them for this emotional perception task. The workers were trained by first thoroughly explaining the emotional attributes of valence, arousal, and dominance. Then, the evaluators were asked to annotate nine conversations from the SEMAINE \cite{McKeown_2012} database. The researcher used four metrics available in CARMA to assess the inter-evaluator agreements between the current worker and previous ``trusted'' annotators. The metrics include the single rater agreement, average rater agreement, single rater consistency, and average rater consistency intraclass correlation coefficients. The researcher then went over each annotation with the worker to discuss any concerns. If the researcher was satisfied after the training session, the worker was allowed to start annotating the actual corpus. The workers remotely annotated the conversation parts on their computers, in their space.

As emotional annotation is an intensive cognitive task, workers were asked to annotate for a maximum of 1 hour before taking a break. The conversation parts to be annotated are presented to the workers in random order, but the emotional attribute to be annotated is chosen in a cycle of 10 consecutive annotations. For example, a worker annotates 10 random conversation parts for valence, and then 10 random conversation parts for arousal. The workers were allowed to annotate a conversation part for an emotional attribute only once. Therefore, if a conversation has six valence annotations, each annotation was made by a different worker.

We assessed annotation quality by calculating inter-annotator agreement and plotting selected annotation traces. We found that one worker consistently produced noisy annotations, and another had low agreement with the other workers. The annotations by the two workers were removed, and their annotations were replaced to achieve a minimum of six annotations.

\subsection{Post-Processing of Annotations}

During annotation, the CARMA software \cite{Girard_2014} samples the location of the joystick at inconsistent times. Furthermore, this inconsistent sampling was dependent on the system CARMA was run in. Except for a few sessions conducted in the laboratory during the project’s initial stages, the annotation process was conducted on the annotators’ personal systems. The raw annotations have an average sampling rate of 33 samples per second. We utilize a median filter to get mean annotations at 10 samples per second. We first pass the individual annotations through a median filter with a window of 120 msec and a shift of 100 msec. Since the maximum distance between samples is more than 120 msec, some windows will be empty during re-sampling. We deal with these empty windows by either repeating the previous sample or setting the value as $0$ if there are no previous samples. We can then average across annotators to get our final emotional traces. 

Time-continuous human annotations exhibit a mismatch between the audio timestamps and the provided annotation values. This occurs because human annotators have a \emph{reaction lag}, defined as the time required for the evaluators to listen to the audio, perceive the emotional content, and move the joystick to the correct position. This reaction time can be compensated for by shifting the final emotional traces by a given reaction lag $T$. We accomplish this by shifting the timing of the trace $T$ sec back, discarding the trace before zero seconds, and padding the trace with its final value to match the audio length. Following Mariooryad and Busso \cite{Mariooryad_2015}, we use a constant reaction lag of $T=3$ secs for all three attributes. As demonstrated in that study, this fixed reaction lag effectively compensates for most of the differences, thereby aligning the emotional behaviors with the emotional traces.

\subsection{Emotional Content of the Corpus}
\label{ssec:emo_cont}

\begin{figure}[t]
    \centering
    \subfloat[Valence]{
        \includegraphics[width=0.37\columnwidth]{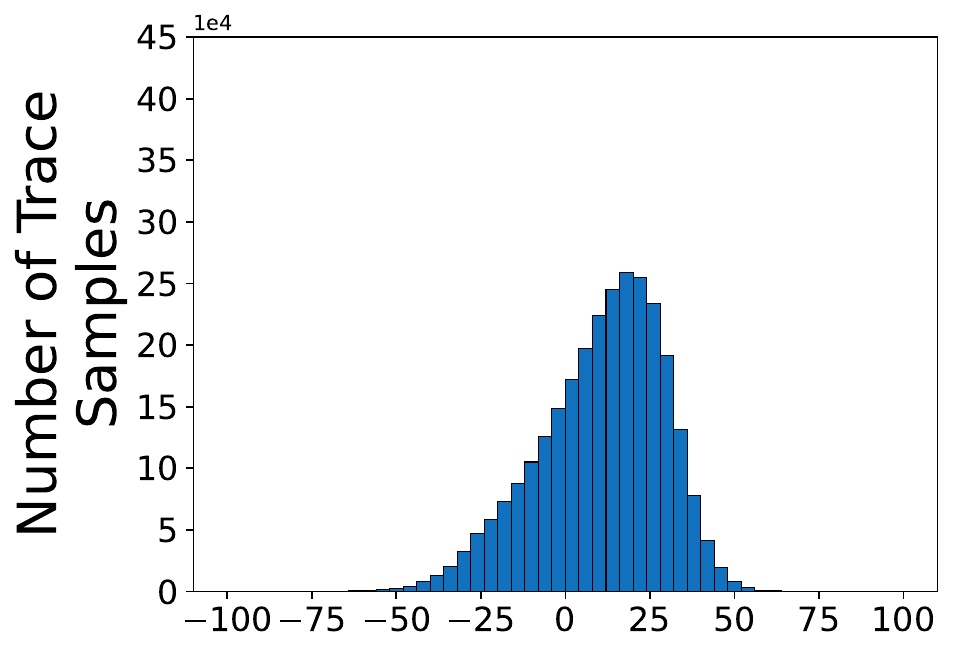}
        \label{fig:emo-cont-val}
    }\hspace{-0.4cm}
    \subfloat[Arousal]{
        \includegraphics[width=0.305\columnwidth]{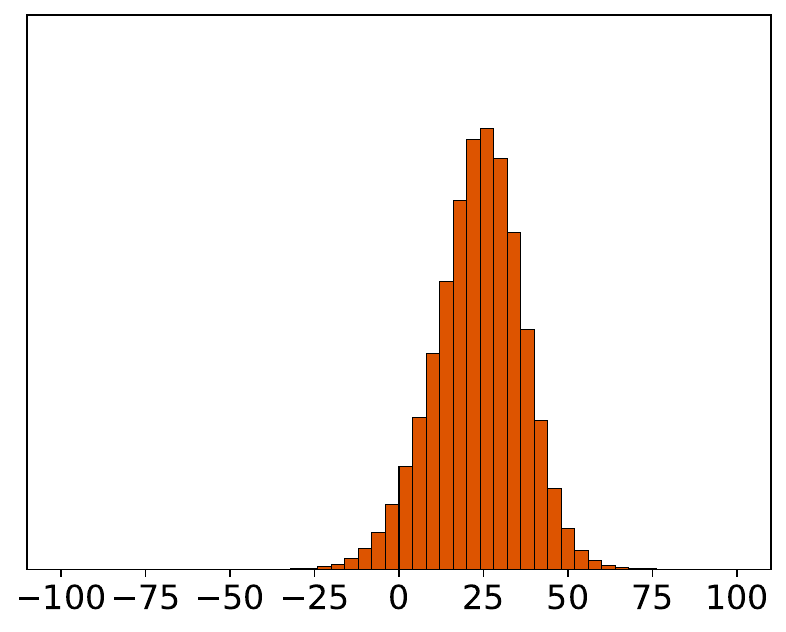}
        \label{fig:emo-cont-aro}
    }\hspace{-0.4cm}
    \subfloat[Dominance]{
        \includegraphics[width=0.305\columnwidth]{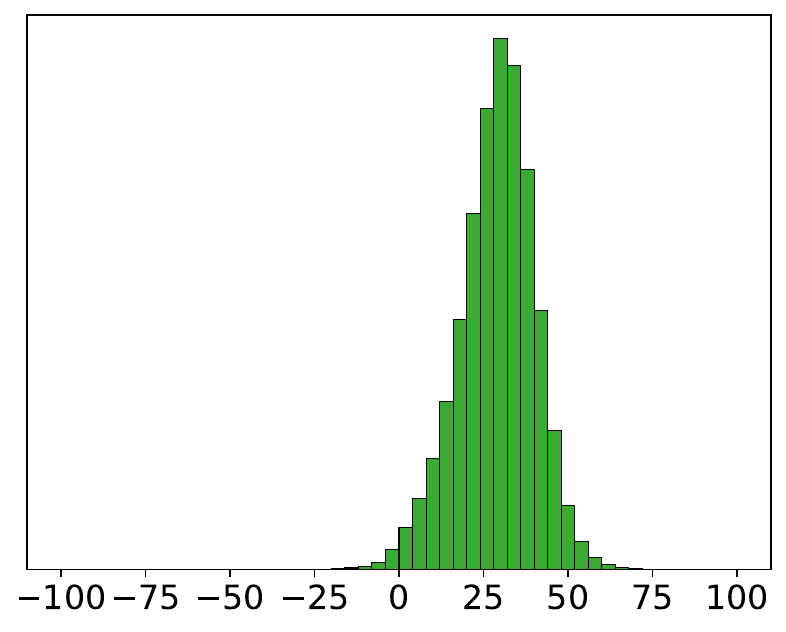}
        \label{fig:emo-cont-dom}
    }
    \caption{Histograms of the attribute values of the mean emotional trace samples, showing the emotional content of the MSP-Conversation corpus.}
    \label{fig:emo-cont}
\end{figure}

We visualize the emotional content of our corpus using the processed annotations. For each attribute, we calculate the mean trace and plot histograms of the frequency of attribute values in these traces. Figure \ref{fig:emo-cont} shows the histograms for all three emotional attributes. The mean of each attribute gives us the typical bias of the corpus: $10.7$ for valence (slightly positive), $23.6$ for arousal (slightly active), and $29.1$ for dominance (slightly strong). The corpus has a positive bias for all three attributes, and valence shows the lowest bias while dominance shows the greatest. The spread of the plots (Fig. \ref{fig:emo-cont}) shows the typical emotional range of the conversations. The standard deviation of the attributes’ values are $18.3$ for valence, $12.9$ for arousal, and $11.0$ for dominance. The plots and statistics show that the corpus has the widest emotional range for valence (Fig. \ref{fig:emo-cont-val}) and the narrowest for dominance (Fig. \ref{fig:emo-cont-dom}).

\subsection{Inter-Evaluator Agreements}
\label{ssec:agreements}

\begin{table}[t]
    \centering
    \caption{Cronbach's Alpha for estimating agreement between annotators of the different data partitions.}
    \begin{tabular}{c|c|c|c}
        \hline
        \textbf{Corpus} & \multicolumn{3}{c}{\textbf{Emotional Attribute}} \\
        \cline{2-4}
        \textbf{Partition} & \textbf{\emph{Valence}} & \textbf{\emph{Arousal}} & \textbf{\emph{Dominance}} \\
        \hline
        \hline
        \emph{Train} & 0.832 & 0.738 & 0.740 \\
        \emph{Development} & 0.832 & 0.719 & 0.656 \\
        \emph{Test1} & 0.821 & 0.705 & 0.673 \\
        \emph{Test2} & 0.736 & 0.633 & 0.543 \\
        \hline
        \emph{All} & 0.849 & 0.758 & 0.745 \\
        \hline
    \end{tabular}
    \label{tab:agreement}
\end{table}

To measure the reliability of our annotations, we estimate the inter-evaluator agreements of our corpus using Cronbach's Alpha \cite{Cronbach_1951}. Cronbach's Alpha measures internal consistency of tests, and we use it to measure the consistency between annotators. An alpha closer to $1$ means that the annotators agree on the ordering of their ratings (i.e., agree on trends). We compute the inter-evaluator agreements between the processed annotations (i.e., traces) of the annotators. Cronbach's alpha can be calculated using the formula:

\begin{equation*}
    \alpha = \frac{k\bar{c}}{\bar{v}+(k-1)\bar{c}}.
\end{equation*}

\noindent
$k$ is the number of samples. For example, for a five-minute conversation part, $k=10\text{ samples/sec}*300\text{ sec}=3,000$. The variable $\bar{c}$ is the mean covariance between annotators, and $\bar{v}$ is the mean annotator variance. Table \ref{tab:agreement} shows the Cronbach's Alpha values of the different data partitions (Sec. \ref{ssec:data_sets}) and the full corpus. The full corpus has alpha values above $0.7$ for all three attributes, showing good agreement. Valence annotations exhibit the highest overall agreement. The dominance annotations show the lowest agreement except in the \emph{Train} partition, where the arousal annotations have the lowest agreement. The high valence alpha values are expected, since valence often has the highest inter-evaluator agreements across attributes \cite{McKeown_2012,Busso_2008_5,Lotfian_2019_3}. Furthermore, during time-continuous annotations, evaluators are able to use context from what they have already heard when rating each subsequent frame. Lee \etal \cite{Lee_2009_2} previously observed that contextual information is especially helpful for the attribute of valence. The low agreement for the attribute of dominance is also expected. The definition of dominance is often more challenging for evaluators to comprehend compared to valence and arousal, resulting in poorer inter-evaluator agreements.

\begin{figure*}[t]
    \centering
    \subfloat[Valence]{
        \includegraphics[width=0.367\linewidth]{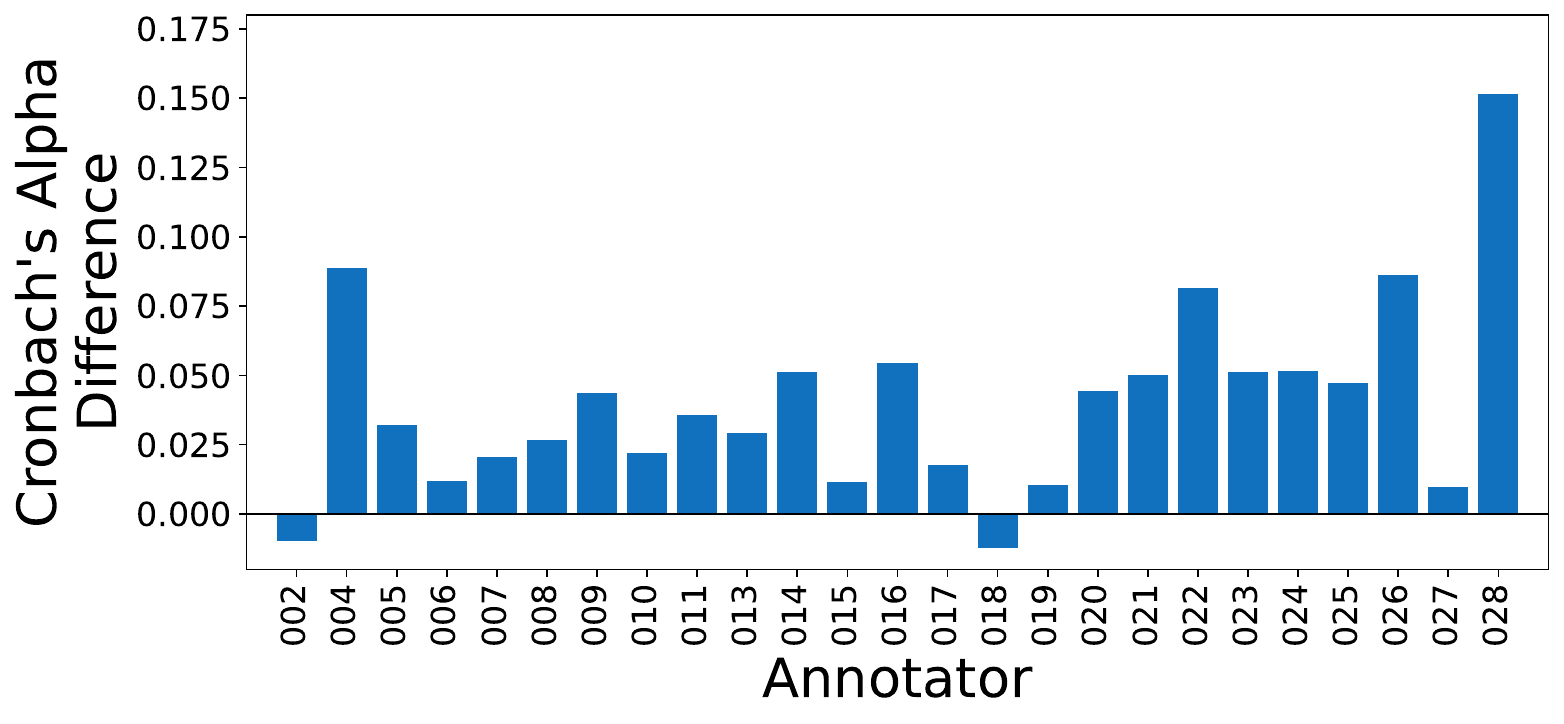}
        \label{fig:agreement_diff_val}
    }\hspace{-0.41cm}
    \subfloat[Arousal]{
        \includegraphics[width=0.3124\linewidth]{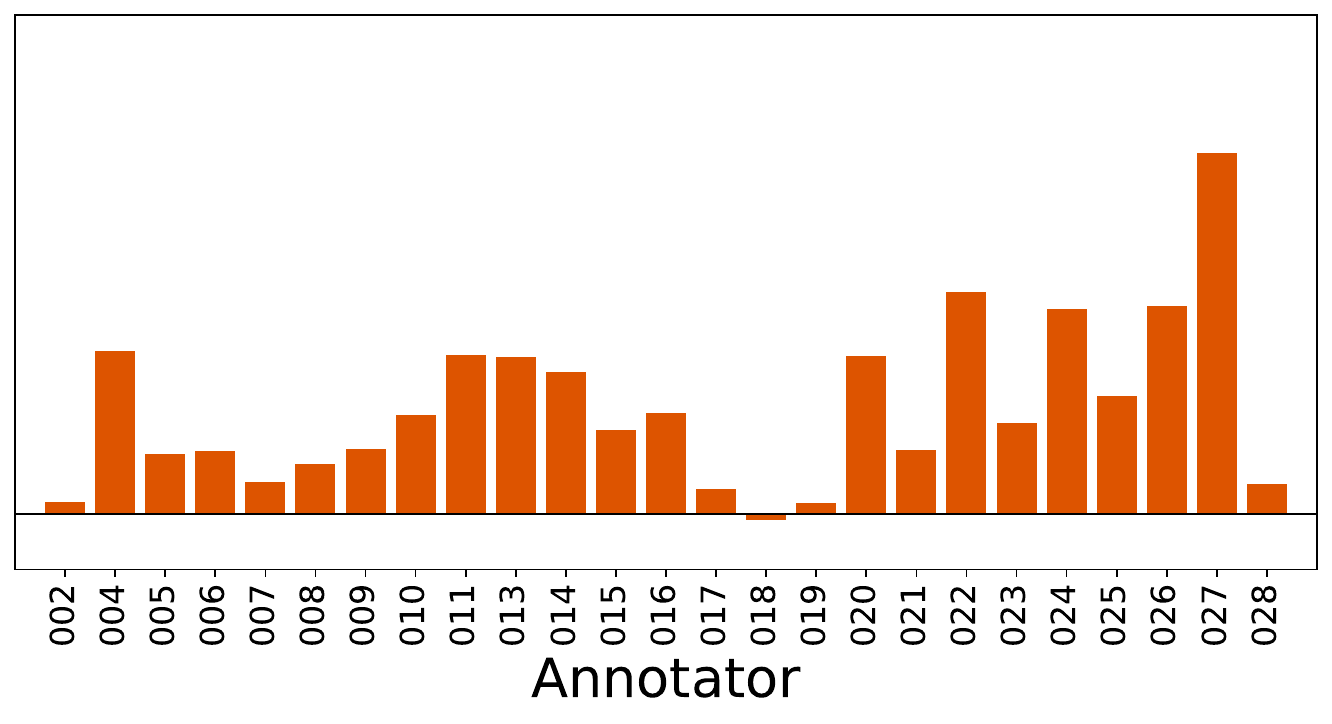}
        \label{fig:agreement_diff_aro}
    }\hspace{-0.41cm}
    \subfloat[Dominance]{
        \includegraphics[width=0.3124\linewidth]{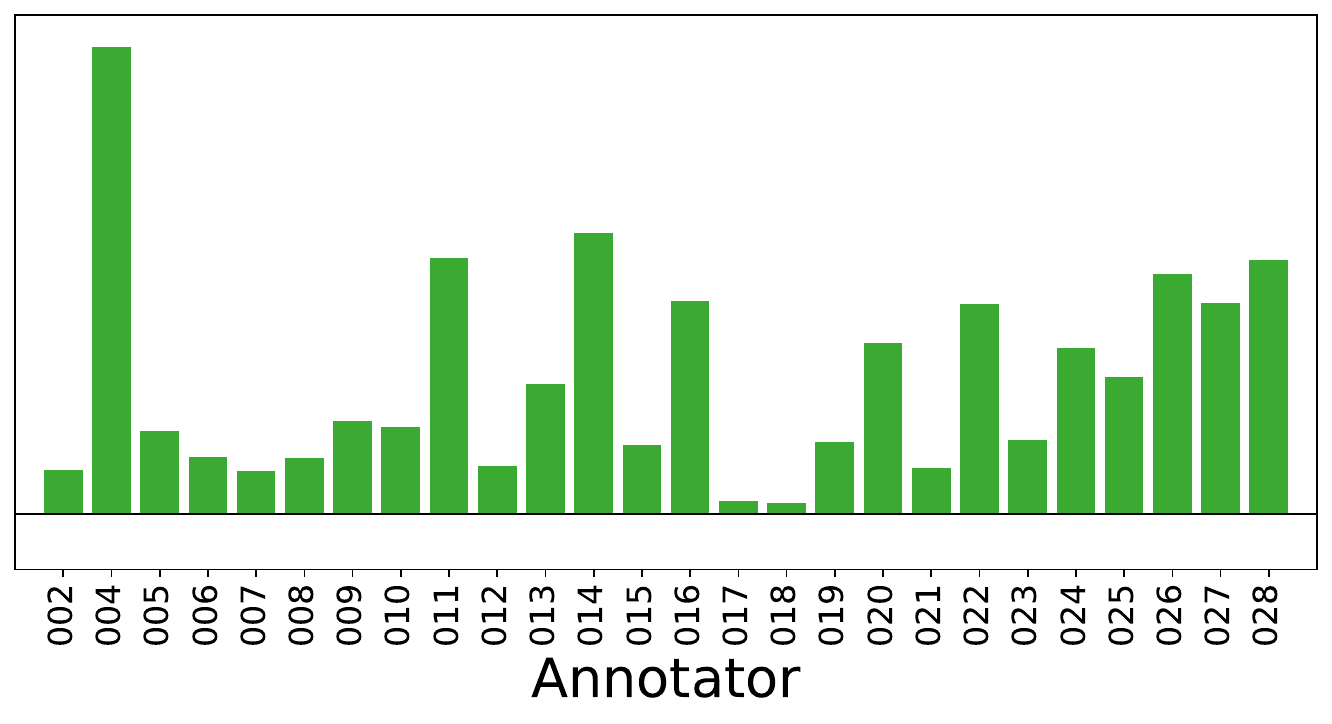}
        \label{fig:agreement_diff_dom}
    }
    \caption{Inter-evaluator agreement differences between excluding then including each annotator. A higher difference value correlates with a more reliable rater. We use Cronbach's Alpha to estimate the agreements between annotators.}
    \label{fig:agreement_diff}
\end{figure*}

We want to evaluate the reliability of each annotator in the corpus. We follow a similar process used for a previous version of the corpus \cite{Martinez-Lucas_2020}. For each annotator $a$, we first collect all conversation parts they annotated. Then, we calculate the inter-evaluator agreement of the traces for those conversation parts. This value is the agreement including the annotator ($\alpha^a_{incl}$). We then remove the traces of the current annotator and again calculate the agreement of the remaining traces. This value is the agreement excluding the annotator ($\alpha^a_{excl}$). We then estimate the agreement difference between excluding and including the annotator $\alpha^a_{diff}=\alpha^a_{incl}-\alpha^a_{excl}$. This difference is positive if including the annotator improves inter-evaluator agreement and negative if it hurts agreement. We again use Cronbach's Alpha to calculate the inter-evaluator agreements. Figure \ref{fig:agreement_diff} shows bar graphs of the inter-evaluator agreement differences of each attribute for each annotator included in the corpus. For valence (Fig. \ref{fig:agreement_diff_val}), only two annotators show a negative agreement difference: annotators $002$ and $018$. For arousal (Fig. \ref{fig:agreement_diff_aro}), only annotator $018$ shows a negative agreement difference. None of the annotators show a negative agreement difference for dominance (Fig. \ref{fig:agreement_diff_dom}). The agreement differences of each annotator can also tell us about the reliability of each annotator. In Sec. \ref{sec:SER}, we use the agreement differences to calculate weighted mean traces to use as labels that are hopefully more reliable.


\section{Partitions of Corpus}
\label{sec:partitions}

The MSP-Conversation corpus contains pre-defined data partitions for use in the development of SER models. Our aim is to have speaker-independent partitions. The MSP-Conversation corpus does not directly include speaker identities. However, as mentioned in Sec. \ref{sec:DataCollection}, all our audio data is sourced from podcasts in the MSP-Podcast dataset \cite{Busso_2025}. The MSP-Podcast corpus contains annotations of speaker identity for its speaking turns, which we can use to ensure speaker-independent partitions. We first look at the MSP-Podcast speaking turns that overlap with the MSP-Conversation corpus to contextualize our strategy to define the partitions in this corpus.

\subsection{Intersection between MSP-Conversation and MSP-Podcast}
\label{ssec:msp-pod_overlap}

The audio data that overlaps between the MSP-Conversation and MSP-Podcast corpora allow us to ask various questions about the two types of annotations employed by the datasets. The MSP-Podcast annotations are completed by evaluators who listen to a single speaking turn and assign a single emotional value to each emotional attribute \cite{Busso_2025}. We call the MSP-Podcast-style annotations \emph{sentence-level} annotations. We call the MSP-Conversation-style annotations (Sec. \ref{sec:Annotation}) \emph{time-continuous} annotations. Table \ref{tab:corr_coeff} shows that 12,555 speaking turns in the MSP-Podcast corpus are included in the recordings of the MSP-Conversation corpus. Our previous work explored the similarities and differences of both types of annotations using previous versions of the MSP-Podcast and MSP-Conversation corpora \cite{Martinez-Lucas_2024_2}. We replicate some of the experiments from our previous work using the most recent versions of both corpora (version 2.0).

\begin{figure}[t]
    \centerline{\includegraphics[width=0.9\columnwidth]{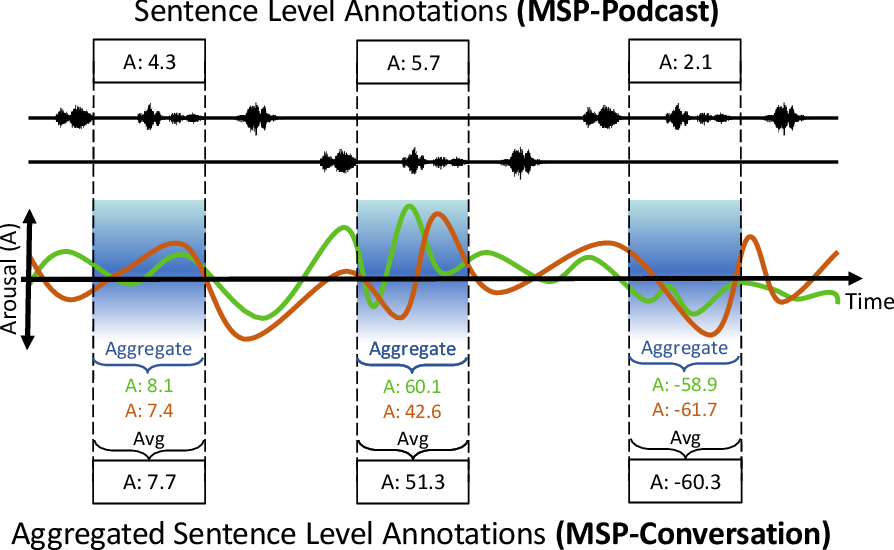}}
    \caption{Process of deriving sentence-level labels analogous to MSP-Podcast labels from the emotional traces in the MSP-Conversation corpus. The figure shows an example using the \emph{arousal} traces of two workers (traces shown in green and orange). The process starts with obtaining the timing of the MSP-Podcast speaking turns overlapping with the current conversation. The trace values within that timing are aggregated using some \emph{aggregation} function. Then the aggregated values are averaged to obtain the sentence-level labels.}
    \label{fig:CTtoSL}
\end{figure}

The first step in this analysis is to convert the MSP-Conversation time-continuous traces into sentence-level labels for comparison with the MSP-Podcast labels. Figure \ref{fig:CTtoSL} shows the process of getting the MSP-Conversation sentence-level labels, which we call the \emph{aggregated} sentence-level labels. The process starts with obtaining the timing of the MSP-Podcast speaking turns with respect to the MSP-Conversation conversation parts that are annotated. The MSP-Conversation annotator traces are then split according to the speaking turn timings. The trace values within each speaking turn are aggregated using an aggregation function. In this work, we use the \emph{mean} as our aggregation function, as it worked well in our previous work \cite{Martinez-Lucas_2024_2}. This strategy gives each speaking turn an aggregated value for each annotator and each attribute. The annotator's aggregated values are then averaged to get an MSP-Conversation sentence-level label of each attribute for each speaking turn. 

\begin{table}[t]
    \centering
    \caption{Pearson Correlation Coefficients between MSP-Podcast labels and MSP-Conversation aggregated sentence-level labels of the same speaking turns for the different MSP-Conversation data partitions.}
    \begin{tabular}{c|c|c|c|c}
        \hline
        \textbf{Corpus} & \textbf{Number of} & \multicolumn{3}{c}{\textbf{Emotional Attribute}} \\
        \cline{3-5}
        \textbf{Partition} & \textbf{Spk Turns} & \textbf{\emph{Valence}} & \textbf{\emph{Arousal}} & \textbf{\emph{Dominance}} \\
        \hline
        \hline
        \emph{Train} & 7,533 & 0.424 & 0.540 & 0.472 \\
        \emph{Development} & 1,570 & 0.521 & 0.554 & 0.391 \\
        \emph{Test1} & 2,718 & 0.502 & 0.502 & 0.369 \\
        \emph{Test2} & 734 & 0.539 & 0.450 & 0.400 \\
        \hline
        \emph{All} & 12,555 & 0.461 & 0.530 & 0.436 \\
        \hline
    \end{tabular}
    \label{tab:corr_coeff}
\end{table}

We want to quantify how similar the two types of annotations are. Since the two corpora have different attribute scales (1 to 7 for MSP-Podcast and -100 to 100 for MSP-Conversation) and different levels of corpus-level biases (see Sec. \ref{ssec:emo_cont}), we use the Pearson Correlation Coefficient to measure the similarity. Table \ref{tab:corr_coeff} shows the Pearson Correlation Coefficients between the MSP-Podcast sentence-level labels and the MSP-Conversation aggregated sentence-level labels for all three attributes and for each MSP-Conversation data partition (Sec. \ref{ssec:data_sets}). Overall, the arousal labels show the highest similarity and the dominance labels the lowest. These results agree with the trends reported in our previous work \cite{Martinez-Lucas_2024_2}. We see that the two annotation types show some correlation, but there are differences that could be important for some applications. For a deeper analysis on the differences between the two types of annotations, please see our previous work \cite{Martinez-Lucas_2024_2}.

\subsection{Training, Development, and Testing Sets}
\label{ssec:data_sets}

\begin{table}
    \centering
    \caption{MSP-Conversation data partitions.}
    \begin{tabular}{c|c|c|c}
        \hline
        \textbf{Corpus} & \textbf{Number of} & \textbf{Number of} & \multirow{2}{*}{\textbf{Duration}} \\
        \textbf{Partition} & \textbf{Conversations} & \textbf{Conv. Parts} & \\
        \hline
        \emph{Train} & 169 & 484 & 41hrs 22min (53.4\%) \\
        \emph{Development} & 52 & 147 & 12hrs 36min (16.3\%) \\
        \emph{Test1} & 58 & 168 & 14hrs 17min (18.5\%) \\
        \emph{Test2} & 31 & 109 & 9hrs 9min (11.8\%) \\
        \hline
        \emph{All} & 310 & 908 & 77hrs 26min (100\%) \\
        \hline
    \end{tabular}
    \label{tab:partitions}
\end{table}

To make speaker-independent partitions of our data, we consider the overlapping speaking turns in the MSP-Podcast corpus. Most overlapping speaking turns have MSP-Podcast speaker identity labels. We also use podcast source metadata, such as the podcast or channel name, to ensure speaker independence in the partitions. We create four data partitions and ensure no MSP-Podcast speaker identity or podcast metadata is common across partitions. 

The partitions and their durations are shown in Table \ref{tab:partitions}. The \emph{Train} partition is the largest and intended to be used to train models. The \emph{Development} partition is intended for hyperparameter tuning during training. The \emph{Test1} partition is intended for evaluating the SER models and providing an estimate of performance on unseen data. Finally, the \emph{Test2} partition is the smallest and intended for fair evaluation across research sets. The emotional labels of the \emph{Test2} partition are not included in any release of the MSP-Conversation corpus. We aim to provide access to a website (https://lab-msp.com/MSP-Conversation\_Competition/DynamicSERB/) where researchers can upload their results on the \emph{Test2} set. This partition allows researchers to compare their performance with other  models submitted by the community.

\section{Speaker Diarizations}
\label{sec:SpkDiarizations}

\begin{figure}[t]
  \centering
  \includegraphics[width=0.9\columnwidth]{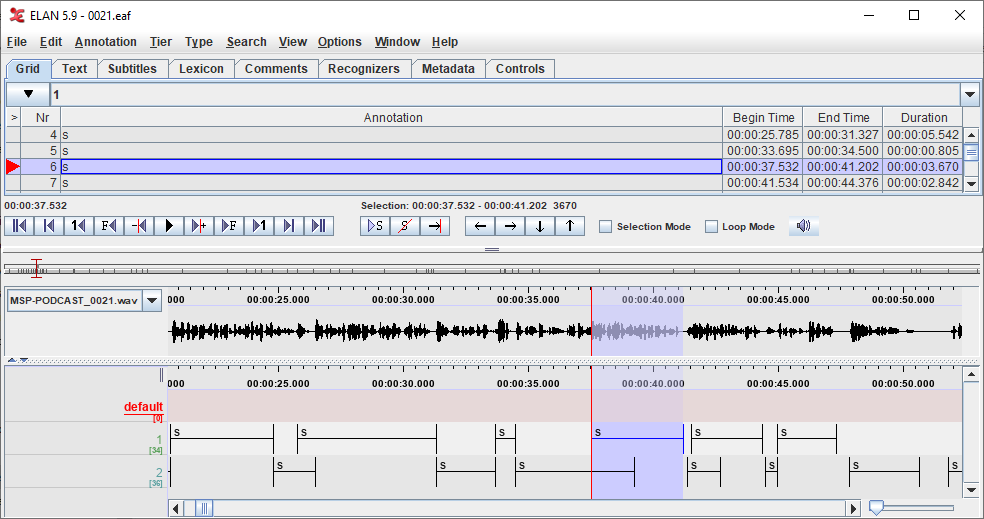}
  \caption{Graphic user interface of ELAN software during the speaker diarization the \emph{MSP-Conversation\_0021} conversation.}
  \label{fig:elan}
\end{figure}

One of the goals in collecting the MSP-Conversation corpus was to provide a resource for conversational speech. To support this goal, we also provide speaker diarizations of all our data. The diarizations are completed by human annotators using the ELAN software \cite{Wittenburg_2006}. Figure \ref{fig:elan} shows an example for a short segment with a completed speaker diarization annotation. Human annotators were instructed to annotate any speech or sound identifiable as specific speakers, including audio such as laughter, crying, and interjections, as well as speech from multiple speakers. Each speaker in a conversation was annotated in a different tier in ELAN, allowing annotators to overlap speech annotations between speakers. During the collection of the corpus, we started providing machine-predicted speaker diarizations as initial \emph{guides} for the annotation process. These guides were obtained using Pyannote \cite{Bredin_2020,Bredin_2021}. The first 89 conversations were fully manually diarized, and the last 221 conversations were diarized starting with the predicted guide.

\begin{figure}[t]
    \centering
    \includegraphics[trim={0 0.9cm 0 0.9cm},clip,width=0.7\columnwidth]{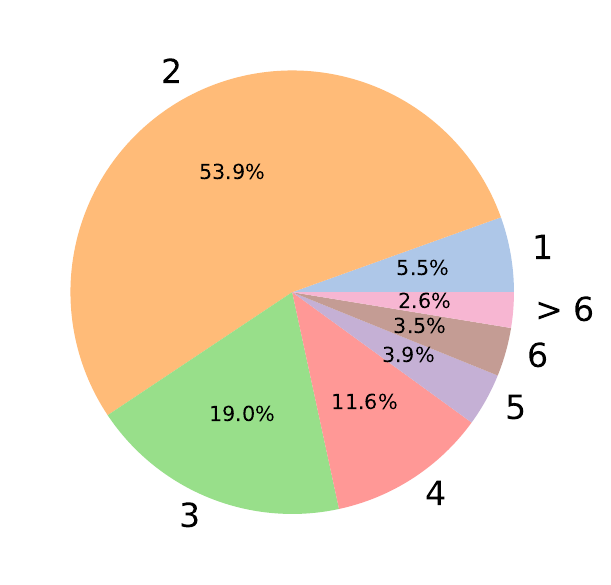}
    \caption{Conversations categorized by the number of speakers identified in their diarization. The majority of the conversations have two or three speakers.}
    \label{fig:pie_spk}
\end{figure}

\begin{table}[t]
    \centering
    \caption{Amount of overlapped speech and non-speech audio.}
    \begin{tabular}{c|c|c}
        \hline
        \textbf{Corpus Partition} & \textbf{Overlapped Speech} & \textbf{Non-Speech Audio} \\
        \hline
        \emph{Train} & 1hr 34min (4.2\%) & 3hrs 27min (8.4\%) \\
        \emph{Development} & 19min (2.7\%) & 54min (7.2\%) \\
        \emph{Test1} & 43min (5.5\%) & 1hr 19min (9.2\%) \\
        \emph{Test2} & 49min (9.2\%) & 18min (3.2\%) \\
        \hline
        \emph{All} & 3hrs 25min (4.8\%) & 5hrs 58min (7.7\%) \\
        \hline
    \end{tabular}  
    \label{tab:over_noise}
\end{table}

Using our speaker diarizations, we can determine the number of speakers in each conversation. Figure \ref{fig:pie_spk} shows a pie chart of the conversations categorized by the number of speakers present. Less than 6\% of the data consists of a single speaker, i.e., monologues. The rest of the data are conversations between two or more speakers. Our corpus also includes a large number of conversations among more than two speakers, accounting for 40.6\% of the data. 

An important aspect of conversations, especially with an increasing number of speakers, is overlapped speech. We calculate the amount of diarized speech that overlaps among two or more speakers. Table \ref{tab:over_noise} shows the duration and percentage of overlapped speech in the entire corpus, as well as in each data partition. Table \ref{tab:partitions} reports the total duration per partition. Table \ref{tab:over_noise} also includes the duration and percentage of the audio which is diarized as coming from none of the speakers, which we call \emph{non-speech audio} (e.g., silence, crowd cheering or laughing, music, noise or distorted audio).

\begin{table}[t]
    \centering
    \caption{Gender of speakers and speech in corpus.}
    \begin{tabular}{c|c|c|c|c|c|c}
        \hline
        \textbf{Corpus} & \multicolumn{3}{c}{\textbf{Number of Speakers}} & \multicolumn{3}{|c}{\textbf{Amount of Speech}} \\
        \cline{2-7}
        \textbf{Partition} & \textbf{\emph{Fem.}} & \textbf{\emph{Male}} & \textbf{\emph{Unk.}} & \textbf{\emph{Fem.}} & \textbf{\emph{Male}} & \textbf{\emph{Unk.}} \\
        \hline
        \hline
        \emph{Train} & 119 & 155 & 80 & 46.0\% & 50.1\% & 5.4\% \\
        \emph{Development} & 39 & 38 & 73 & 51.8\% & 31.8\% & 17.7\% \\
        \emph{Test1} & 35 & 28 & 12 & 66.9\% & 28.4\% & 5.6\% \\
        \emph{Test2} & 18 & 26 & 11 & 42.8\% & 52.3\% & 6.9\% \\
        \hline
        \emph{All} & 211 & 247 & 176 & 50.4\% & 43.4\% & 7.6\% \\
        \hline
    \end{tabular}  
    \label{tab:gender}
\end{table}

We annotate the first 74 conversations with each speaker's apparent gender. To obtain the speakers’ genders in the remaining conversations, we use the overlapped MSP-Podcast speaker turns (Sec. \ref{ssec:msp-pod_overlap}). Using our diarizations, we match each overlapped speaker turn to a speaker in the corresponding conversation. We then use the MSP-Podcast speaker labels to match the most likely MSP-Podcast speaker to each speaker in each conversation. We finally use the MSP-Podcast gender label for each speaker. Some of the MSP-Conversation speakers do not have overlaps with known MSP-Podcast speakers. We set the gender of those speakers as unknown. Table \ref{tab:gender} shows the number of speakers by gender and the amount of speech by gender in the full corpus and in each data partition. The full corpus shows a bias toward female speech, although we have identified more male speakers. There are some variations across partitions in the number of speakers and total speech by gender. The Train partition is the most balanced set, providing a good basis for training generalizable SER models.

\section{Speech Emotion Recognition Modeling}
\label{sec:SER}

In this section, we build, train, and evaluate dynamic SER models on the MSP-Conversation corpus to predict emotional traces that describe emotional attributes in the conversations. The results shown in this section act as a baseline performance for future work using the MSP-Conversation corpus for SER.

\subsection{Architecture Implementation}
\label{sssec:modeling}

The SER models have three stages. The first stage extracts speech representations from the audio. We use a \emph{self-supervised learning} (SSL) framework to get representation vectors for the audio data. We utilize the transformer-based ``wavlm-large'' architecture \cite{Chen_2022} by first loading the pre-trained ``wavlm-large'' model from the HugginFace library \cite{Wolf_2019}. We then load WavLM attention encoder weights that have been fine-tuned using version 2.0 of the MSP-Podcast corpus \cite{Busso_2025}. The fine-tuning was performed in a multi-task manner, using a WavLM + SER head model to predict the valence, arousal, and dominance values for each speaking turn. We use the output of the WavLM attention encoder as our speech representation vectors.

The second stage further processes the representations for our task. It processes the input feature vectors to lower their dimensionality and focus the representations on our specific task. To match the sample rate of our emotional traces, we use mean-pooling with a window size of 120ms and step size of 100ms on the output vectors from the first stage. We pass the new vectors through a feed-forward neural network. This network consists of three linear layers of size $512$ with \emph{rectified linear unit} (ReLU) activation and dropout at a rate of $0.5$ between each layer. 

The third stage uses those representations to predict the audio’s emotional attributes, creating a SER head. The SER model is expected to predict either one of the attributes per model (single-task), or the three attributes together (multi-task), which has shown to improve SER performance \cite{Naini_2025_3, Parthasarathy_2017_3}. We build three SER heads using either a \emph{Linear} layer, an \emph{LSTM} model, or a \emph{Transformer} module. The \emph{Linear} SER head is a simple four-layer linear neural network. The first three layers have size $512$, and the last layer has size $1$ (single-task) or $3$ (multi-task). We also add ReLU activation and dropout at a rate of $0.5$ between them. The \emph{LSTM} SER head is a \emph{bidirectional long short-term memory} (BiLSTM) network. It consists of two BiLSTM layers with hidden size $256$, with dropout at a rate of $0.5$ between them. Then, the output of the last BiLSTM layer is passed through the output linear layer, which has a size of $1$ (singletask) or $3$ (multi-task) depending on the task. The final SER head is the \emph{Transformer} head. We use the Transformer decoder architecture used by Vaswani~\etal \cite{Vaswani_2017}. We use $8$ heads for the multi-head attention and $6$ decoder layers. The output of the Transformer decoder is then passed through the task-specific output linear layer.

\subsection{Training Details}
\label{ssec:training}

We use the processed emotional traces as our labels when training our dynamic SER models. Therefore, we train and evaluate the models on the three- to seven-minute conversation segments used for annotation (Sec. \ref{sec:Annotation}). These conversation parts are on average five minutes long, but the WavLM model is trained on shorter speech utterances \cite{Chen_2022}. To ensure that the WavLM attention encoder can effectively output speech representations, we further split our conversation parts into sections of up to one minute with five-second overlaps. During evaluation, we truncate the one-minute sections at the midpoint of their overlaps and concatenate them to obtain our predicted trace.

We train our proposed model on the training set of the MSP-Conversation corpus. Our loss function is the \emph{concordance correlation coefficient} (CCC) loss:

\begin{equation*}
    \mathcal{L} = 1 - CCC(y,\hat{y}) = 1 - \frac{2\rho_{y,\hat{y}}\sigma_{y}\sigma_{\hat{y}}}{\sigma_{y}^2+\sigma_{\hat{y}}^2+(\mu_y-\mu_{\hat{y}})^2}
\end{equation*}

\noindent
where $\rho_{y,\hat{y}}$ is the Pearson Correlation Coefficient between the label traces $y$ and the predicted traces $\hat{y}$, and $\mu$ and $\sigma^2$ are the mean and variance of the traces. As mentioned in Section \ref{sssec:modeling}, we have two training scenarios: single and multi-task. For single-task training, we predict a single emotional attribute at a time. The loss function is just the loss for the current attribute predictions. For multi-task training, we predict all three emotional attributes. The loss function is then $\mathcal{L}_{MT}=\omega_{val}\mathcal{L}_{val}+\omega_{aro}\mathcal{L}_{aro}+\omega_{dom}\mathcal{L}_{dom}$ where $\omega_{val}+\omega_{aro}+\omega_{dom} = 1$. We use $\omega_{val}=\omega_{aro}=\omega_{dom}=\frac13$ For our baseline experiments.

We train the \emph{Linear} head model for 25 epochs and the \emph{LSTM} and \emph{Transformer} head models for 50 epochs. For all our models, we use the Adam optimizer \cite{Kingma_2014_2} with a learning rate of 0.0001 and a batch size of 64 one-minute sections. We evaluate the models on the Development set at each epoch and choose the best model as the final model. Finally, we evaluate the models on the Test1 and Test2 sets. We use the CCC as our evaluation metric.  A CCC closer to 1 shows better performance.

\subsection{Baseline Results}
\label{ssec:results}

\begin{table}[t]
    \centering
    \caption{Test CCC results of baseline SER models for single-task (\emph{ST}) and multi-task (\emph{MT}) training.}
    \begin{tabular}{cc|c|c|c|c|c|c}
        \hline
         & \multirow{2}{*}{\textbf{SER Head}} & \multicolumn{2}{c}{\textbf{Valence}} & \multicolumn{2}{|c}{\textbf{Arousal}} & \multicolumn{2}{|c}{\textbf{Dominance}} \\
        \cline{3-8}
         & & \textbf{\emph{ST}} & \textbf{\emph{MT}} & \textbf{\emph{ST}} & \textbf{\emph{MT}} & \textbf{\emph{ST}} & \textbf{\emph{MT}} \\
        \hline
        \hline
        \multirow{3}{*}{\rotatebox[origin=c]{90}{Test1}} & \emph{Linear} & 0.671 & 0.674 & 0.619 & 0.576 & 0.457 & 0.489 \\
         & \emph{LSTM} & 0.676 & 0.675 & 0.669 & 0.634 & 0.517 & 0.525 \\
         & \emph{Transformer} & 0.666 & 0.675 & 0.612 & 0.608 & 0.487 & 0.510 \\
        \hline
        \multirow{3}{*}{\rotatebox[origin=c]{90}{Test2}} & \emph{Linear} & 0.582 & 0.563 & 0.486 & 0.413 & 0.367 & 0.412 \\
         & \emph{LSTM} & 0.589 & 0.518 & 0.488 & 0.450 & 0.379 & 0.397 \\
         & \emph{Transformer} & 0.569 & 0.553 & 0.430 & 0.422 & 0.370 & 0.350 \\
        \hline
    \end{tabular}
    \label{tab:ser_base}
\end{table}

Table \ref{tab:ser_base} shows the CCC results on our two test sets. In general, the model with the \emph{LSTM} head performs the best. Although it is more complex, the \emph{Transformer} head performs below the \emph{LSTM} head. The BiLSTM network, being a different structure than the WavLM network, allows the model to encode more information that is not present in the features. Whereas the Transformer decoder, which follows a structure similar to the WavLM attention encoder, does not encode as much novel information. When we compare single-task and multi-task training, we see that single-task is better for arousal and multi-task is better for dominance. The model might gain useful information from the other two attributes when predicting dominance. The extra information is especially useful for predicting dominance in the MSP-Conversation corpus, since the models perform worst on this attribute. Our results also show that predicting Test2 is a harder task than predicting Test1. This result is expected since Test2 has lower inter-evaluator agreement (Table \ref{tab:agreement}), more overlapped speech (Table \ref{tab:over_noise}), and a different gender ratio than the rest of the data (Table \ref{tab:gender}).

\section{Discussion}
\label{sec:discussion}
The MSP-Conversation corpus opens up various research directions given its size, diversity, and naturalness. The large amount of data annotated in a time-continuous manner enables the study of contextual information in emotion perception (Sec. \ref{ssec:context}). The three to seven-minute length of the emotional traces provides annotations of both short and long-term emotional dynamics, giving important data for their study (Sec. \ref{ssec:emo_dynamics}). The overlap between the MSP-Conversation and MSP-Podcast datasets gives an important resource for exploring the use of data in new tasks and environments (Sec. \ref{ssec:overlap_corpora}). Finally, the availability of speaker diarizations of the annotated conversations enables the study of emotions in group conversations (Sec. \ref{ssec:group_conv}).

\subsection{Understanding the Role of Contextual Information}
\label{ssec:context}

The MSP-Conversation corpus contains over 70 hours of speech data annotated in a time-continuous manner. During the annotation process of a conversation part, raters are given an increasing amount of context, defined as the surrounding speech from the source recording, for evaluating the emotions present in the audio. Previous studies have shown that the presence of context changes emotional perceptions and annotations of speech \cite{Cauldwell_2000, Jaiswal_2019}. The increasing context of the MSP-Conversation annotations can be used for further research on the effect of varying amounts of context. The annotation of different conversation parts is done sequentially, one after the other, and the date and time each annotation was started is available. This information opens up research paths into the effects of various types of context on emotional perception.


\subsection{Studying Short and Long-Term Emotional Dynamics}
\label{ssec:emo_dynamics}

The time-continuous emotional annotations in the MSP-Conversation corpus are longer than those in most emotional speech corpora. The length of the conversation varies, with both short- and long-term emotional displays. In the short-term, interesting emotional dynamics include sudden shifts \cite{Huang_2015,Huang_2016}, whose causes and intensities can provide important information for emotional models. Another short-term effect is the presence of emotionally salient regions \cite{Parthasarathy_2016,Parthasarathy_2021}, which are often easier to perceive and predict as they stand out from their surrounding emotional content. The corpus can also help in the study of long-term emotional phenomena such as emotion propagation and emotional contagion \cite{Hatfield_1993}. These phenomena can be described as people ``catching'' the emotions of others as they interact with them, analogous to disease propagation. These emotional dynamics affect interactions between people over longer periods. The emotional traces in the corpus can be used to study and model these dynamics.

\subsection{Exploring the Overlap of Different Corpora}
\label{ssec:overlap_corpora}

A unique and useful resource provided by the MSP-Conversation corpus is its overlap with the MSP-Podcast corpus (Sec. \ref{ssec:msp-pod_overlap}). Our previous work explored the validity of aggregating (Fig. \ref{fig:CTtoSL}) time-continuous annotations over short speaking turns that can be used to train common sentence-level SER models \cite{Martinez-Lucas_2024_2}. As reliable emotional annotations are difficult to obtain, maximizing the amount of emotional data available to train modern models is crucial. Another research direction that can be explored by the overlap between the datasets is the creation of pseudo-labels of emotional aspects not originally annotated. For example, the MSP-Podcast corpus contains attribute and categorical emotional labels, while the MSP-Conversation only has attribute annotations. Using the overlap between the corpora, categorical pseudo-labels based on the MSP-Podcast corpus could be created for the MSP-Conversation corpus.

Another relevant database is the NaturalVoices corpus \cite{Du_2025,Salman_2024}, which also relies on the podcast recording used for the MSP-Conversation corpus, focusing on speech generation tasks \cite{Ulgen_2024_2,Mahapatra_2025}.

\subsection{Analyzing Group Conversations of Varying Sizes}
\label{ssec:group_conv}

All the conversations in the MSP-Conversation corpus are manually diarized for speaker activity. This information makes the corpus an ideal resource for studying the emotional dynamics between speakers. Further, Figure \ref{fig:pie_spk} shows that 40.6\% of the conversations are between more than two speakers. The figure also shows the diversity of group sizes for the conversations. These aspects of the dataset make it a valuable resource for studying both small and large groups. Conversations between two speakers are easier to track as there are fewer states. For example, if we ask who is speaking or who a speaker is responding or speaking to, a model only has four choices in a two-speaker conversation (speaker 1, speaker 2, both, or none). The larger the group in a conversation, the higher the likelihood of confusion or mixed emotions that are harder to predict. However, larger group conversations are common in human interactions, so their study is necessary for naturalistic affective computing.

\section{Conclusions}
\label{sec:conclusions}

This paper presented the development and analysis of the MSP-Conversation dataset. Our proposed natural speech corpus contains interactions among varying numbers of speakers, rich emotional annotations across three emotional attributes, and human speaker diarizations. Our emotional annotations are completed in a time-continuous manner, enabling further research on the role of contextual information and the study of short and long-term emotional dynamics. To obtain reliable annotations, we provide at least six ratings completed by different raters for each speech sample and attribute. We ensure high-quality annotations by providing in-depth training and continual evaluation to each rater. One of our goals for this corpus was to provide a resource for SER methods that predict time-continuous emotional traces (Dynamic SER). In this paper, we show Dynamic SER modeling results on the proposed MSP-Conversation corpus, providing baselines for further research. 

Although the MSP-Conversation corpus is a great resource for affective computing work with context and dynamics, it has some limitations. The natural speech that comprises the dataset limits the number of clear and extreme annotated emotional events in the corpus compared to datasets with acted or prompted speech. The emotional content in these recordings could bias models if not taken into account or counter-acted. Further, the use of human annotations (emotional ratings and speaker diarizations) comes with the inherent possibility of errors. The use of inter-evaluator agreements to monitor ratings does not fully ensure the accuracy of the labels, and the human diarization of each conversation was completed by a single person. Overall, the corpus focuses on a single modality (speech), and a single language (English). The complexity of emotion expression and perception makes the additional use of video and text an attractive and effective method for improving emotion recognition. The outlining of these limitations gives us directions to improve or augment the MSP-Conversation corpus in the future.

\bibliographystyle{IEEEtran}
\bibliography{references_Martinez-Lucas}


\vspace{-1.3cm}

\begin{IEEEbiography}[{\includegraphics[width=1in,height=1.25in,clip,keepaspectratio]{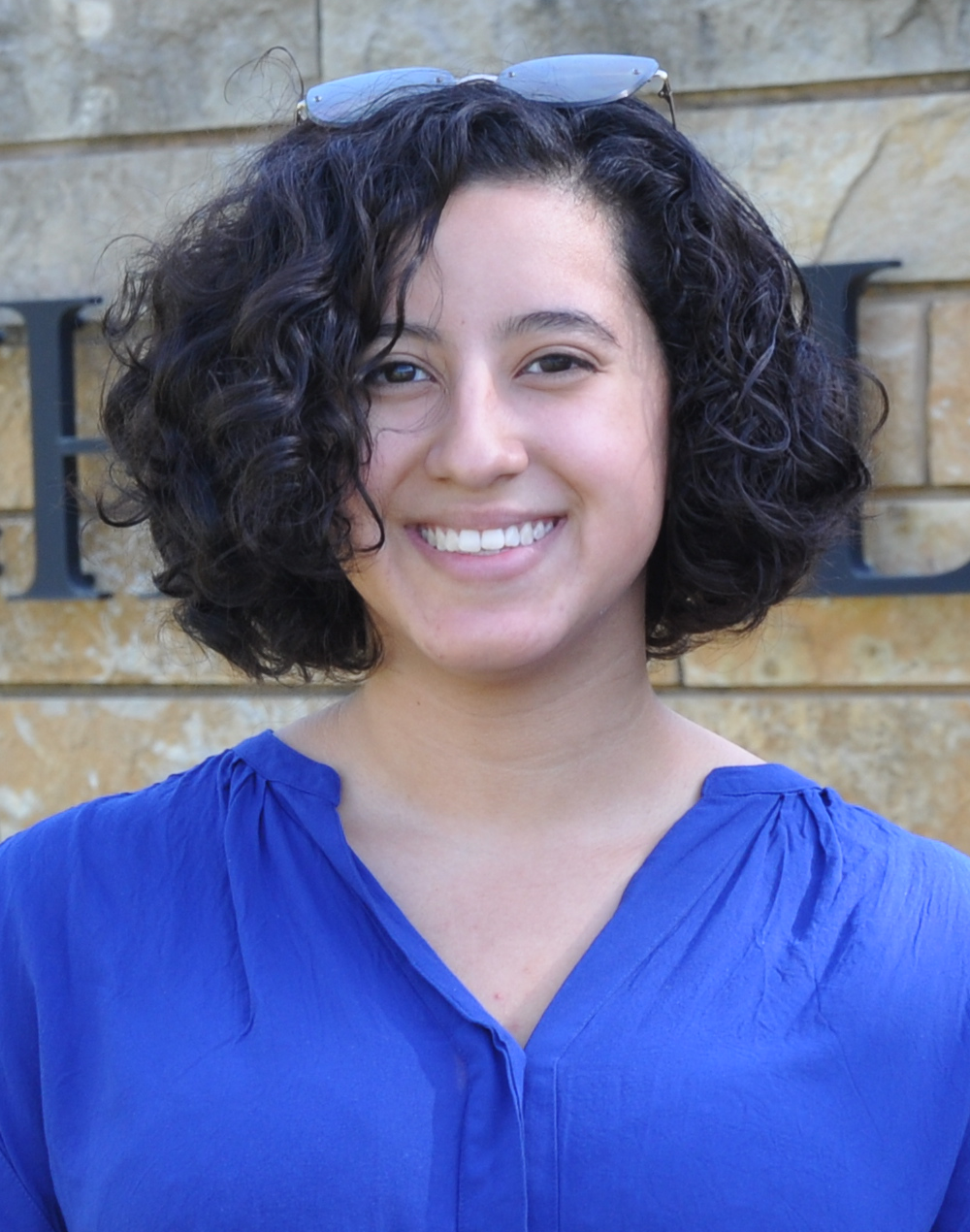}}]{Luz Martinez-Lucas} (S'21) is a PhD Student in the Electrical and Computer Engineering Department at the University of Texas at Dallas (UTD). She did her Bachelor's in Electrical Engineering at UTD. Her research interests include affective computing, speech technology, and machine learning. She is a student member of IEEE and AAAC.
\end{IEEEbiography}

\vspace{-1.0cm}

\begin{IEEEbiography}
[{\includegraphics[width=1in,height=1.25in,clip,keepaspectratio]{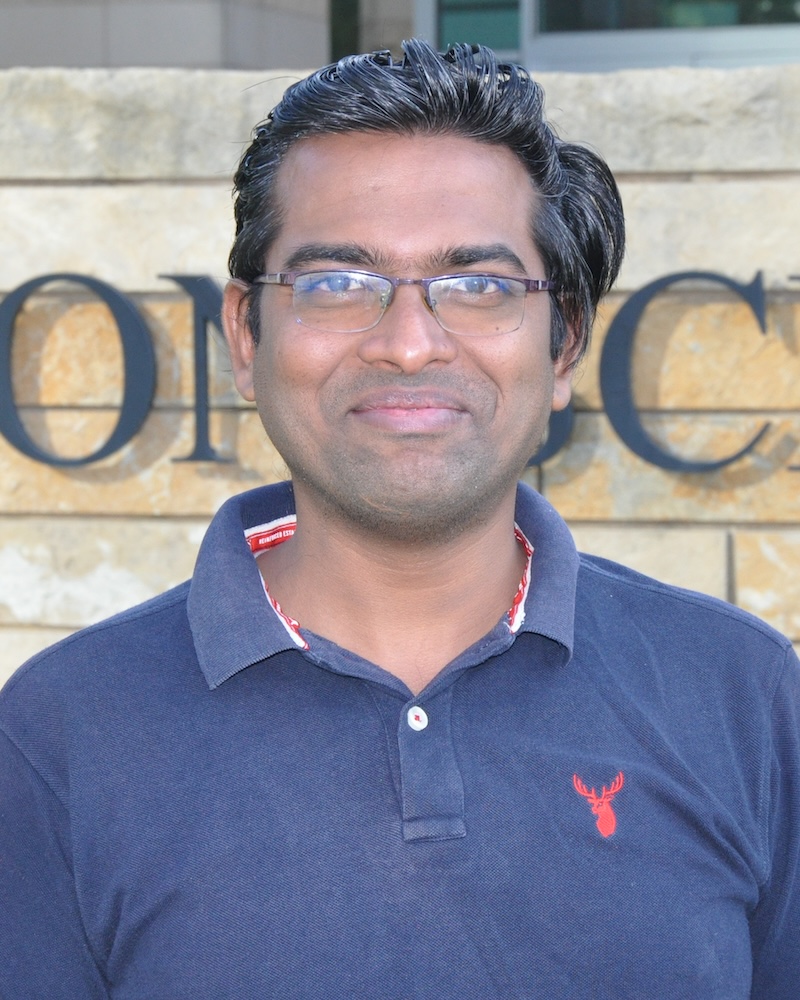}}]{Pravin Mote} (S'25) is currently pursuing a Ph.D. in the Department of Electrical and Computer Engineering at the University of Texas at Dallas. He is also a visiting researcher at the Language Technologies Institute, Carnegie Mellon University. His research interests include speech technology, multimodal affective computing, and machine learning.
\end{IEEEbiography}

\vspace{-1.0cm}

\begin{IEEEbiography}[{\includegraphics[width=1in,height=1.25in,clip,keepaspectratio]{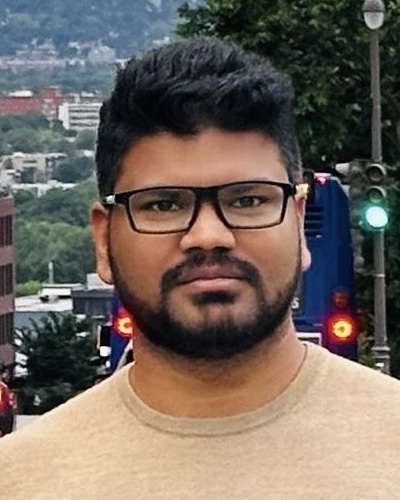}}]{Abinay Reddy Naini} (S'19)  is a PhD student in the Department of Electrical and Computer Engineering at the University of Texas at Dallas (UTD) and is currently working as a visiting researcher at the Language Technologies Institute, Carnegie Mellon University. He received his B.S. in Electrical Engineering from the National Institute of Technology, Warangal, India, and his M.S. in Electrical Engineering from the Indian Institute of Science (IISc). His research interests include affective computing, speech technology, and machine learning. 
\end{IEEEbiography}

\vspace{-1.0cm}

 \begin{IEEEbiography}[{\includegraphics[width=1in,height=1.25in,clip,keepaspectratio]{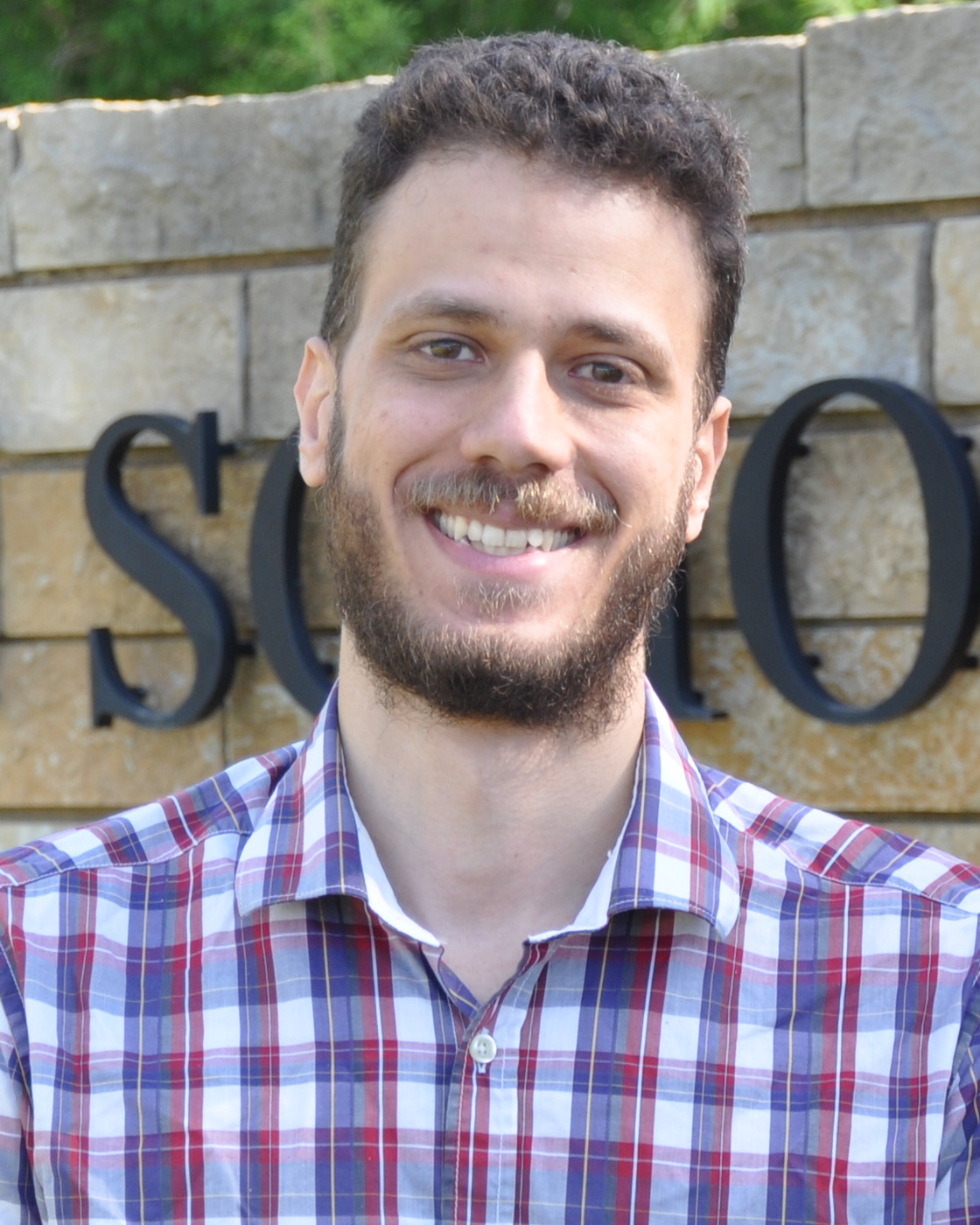}}]{Mohammed AbdelWahab} (S'14) received his B.Sc. degree in electrical and electronic engineering at Ain Shams University, Cairo, Egypt in 2010, and his M.S degree in Electrical engineering from Nile university, Cairo, Egypt 2012. He received his Ph.D. degree in electrical engineering at the University of Texas at Dallas. His current research interest includes speech signal processing, emotion recognition, artificial intelligence and machine learning.
 \end{IEEEbiography}

\vspace{-1.0cm}

\begin{IEEEbiography}[{\includegraphics[width=1in,height=1.25in,clip,keepaspectratio]{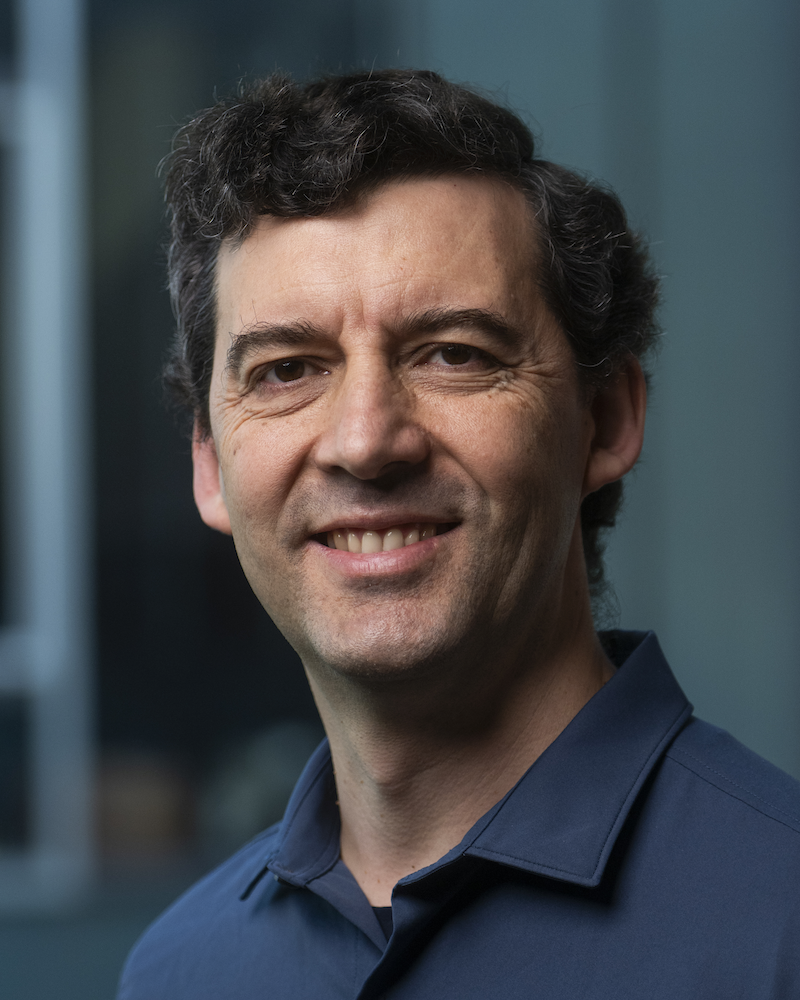}}]{Carlos Busso} (S’02-M'09-SM'13-F'23) is a Professor at Language Technologies Institute, Carnegie Mellon University, where he is also the director of the \emph{Multimodal Speech Processing} (MSP) Laboratory. He received the BS and MS degrees with high honors in electrical engineering from the University of Chile, Santiago, Chile, in 2000 and 2003, respectively, and the PhD degree (2008) in electrical engineering from the University of Southern California (USC), Los Angeles, in 2008. His research interest is in human-centered multimodal machine intelligence and applications, focusing on the broad areas of speech processing, affective computing, multimodal behavior generative models, and foundational models for multimodal processing. He was selected by the School of Engineering of Chile as the best electrical engineer who graduated in 2003 from Chilean universities. He is a recipient of an NSF CAREER Award. In 2014, he received the ICMI Ten-Year Technical Impact Award. His students received the third prize IEEE ITSS Best Dissertation Award (N. Li) in 2015, and the AAAC Student Dissertation Award (W.-C. Lin) in 2024. He also received the Hewlett Packard Best Paper Award at the IEEE ICME 2011 (with J. Jain), and the Best Paper Award at the AAAC ACII 2017 (with Yannakakis and Cowie). He received the Best of IEEE Transactions on Affective Computing Paper Collection in 2021 (with R. Lotfian) and the Best Paper Award from IEEE Transactions on Affective Computing in 2022 (with Yannakakis and Cowie). In 2023, he received the Distinguished Alumni Award in the Mid-Career/Academia category by the \emph{Signal and Image Processing Institute} (SIPI) at the University of Southern California. He received the 2023 ACM ICMI Community Service Award. He is currently a Senior Area Editor of IEEE/ACM Speech and Language Processing.  He is a member of AAAC and a senior member of ACM. He is an IEEE Fellow and an ISCA Fellow.
\end{IEEEbiography}




\end{document}